\documentclass[aps,prb,twocolumn,superscriptaddress,showpacs]{revtex4}

\usepackage{amsmath}
\usepackage{bm}
\usepackage{graphicx}

\begin{document}

\title{Wannier-Stark ladder in the linear absorption of a random system with
scale-free disorder}

\author{E.\ D\'{\i}az}
\thanks{Also at Grupo Interdisciplinar de Sistemas Complejos.}
\author{F.\ Dom\'{\i}nguez-Adame}
\thanks{Also at Grupo Interdisciplinar de Sistemas Complejos.}
\affiliation{Departamento de F\'{\i}sica de Materiales, Universidad Complutense,
E-28040 Madrid, Spain}

\author{Yu.\ A.\ Kosevich}
\email[]{yukosevich@yahoo.com} 
\affiliation{N. N. Semenov Institute of Chemical Physics, Russian Academy of
Sciences, ul. Kosygina 4, 119991 Moscow, Russia} 
\affiliation{Nanophotonics Technology Center, Polytechnic University of
Valencia, C/ Camino de Vera s/n, E-46022 Valencia, Spain}

\author{V.\ A.\ Malyshev}
\thanks{On leave from "S.I. Vavilov State Optical Institute", Birzhevaya Linia
12, 199034 Saint-Petersburg, Russia} 
\affiliation{Institute for Theoretical Physics and Materials Science Centre,
University of Groningen, Nijenborgh 4, 9747 AG Groningen, The Netherlands}

\begin{abstract}

We study numerically the linear optical response of a
quasiparticle  moving on a one-dimensional disordered lattice in
the presence of a linear bias. The random site potential is
assumed to be long-range-correlated with a power-law spectral
density $S(k) \sim 1/k^{\alpha}$, $\alpha > 0$. This type of
correlations results in a phase of extended states at the band
center, provided $\alpha$ is larger than a critical value
$\alpha_c$ [F.\ A.\ B.\ F.\ de Moura and M.\ L.\ Lyra, Phys. Rev.
Lett. \textbf{81}, 3735 (1998)]. The width of the delocalized
phase can be tested by applying an external electric field:
Bloch-like oscillations of a quasiparticle wave packet are
governed by the two mobility edges, playing now the role of band
edges [F.\ Dom\'{\i}nguez-Adame \emph{et al.}, Phys.\ Rev.\ Lett.\
\textbf{91}, 197402 (2003)]. We demonstrate that the
frequency-domain counterpart of these oscillations, the so-called
Wannier-Stark ladder, also arises in this system. When the phase
of extended states emerges in the system, this ladder turns out to
be a comb of doublets, for some range of disorder strength and
bias. Linear optical absorption provides a tool to detect this
level structure.

\end{abstract}


\maketitle

\section{Introduction}

\label{intro}

Since 1998,~\cite{Moura98} there exists an increasing interest in
studying the localization properties of the quasiparticle wave
functions in one-dimensional (1D) disordered systems with a
long-range-correlated site potential
landscape.~\cite{Izrailev99,Liu99,Kantelhardt00,Kuhl00,Russ01,Lindquist01,%
Chen02,Zhang02,Carpena02,Lima02,Moura02,Adame03,Liu03,Carpena04,Moura04,%
Ndawana04,Yamada04,Shima04,Albuquerque05,Cheraghchi05,Russ98}
Random sequences, having a power-law spectral density $S(k) \sim
1/k^{\alpha}$ with $\alpha > 0$, result in a phase of extended
states at the band center, provided $\alpha$ is larger than a
critical value $\alpha_c$.~\cite{Moura98,Shima04} This finding
contradicts the widely admitted conclusion of the one-parameter
scaling theory of localization~\cite{Abrahams79} that all states
of noninteracting quasiparticles in one and two dimensions with
time reversal symmetry are localized (see
Refs.~\onlinecite{Lee85,Kramer93,Beenakker97,Janssen98} for an
overview). As a matter of fact, a great deal of work is being
devoted to put the correlation-induced low-dimensional
localization-delocalization transition (LDT) on solid grounds.

It is worthwhile to notice that long-range correlations with
spectral density of the form $1/k^{\alpha}$ are widely presented
in nature, both in \emph{vitro\/} and in \emph{vivo\/} (see, e.g.,
Refs.~\onlinecite{Paczuski96} and~\onlinecite{Havlin99} and
references therein). This type of correlation gives rise to the
"fractal geometry of nature" introduced by
Mandelbrot.~\cite{Mandelbrot82} Importantly, in 1992 it was
conjectured that the long-range power-law correlations exist in
nucleotide sequences in DNA.~\cite{Peng92,Li92,Voss92,Stanley93}
This conjecture opened prospectives to quantify nature \emph{in
vivo\/} with critical exponents.~\cite{Buldyrev98} It was argued
also that long-range correlations in DNA sequences can explain the
long-distance charge transport in these
systems.~\cite{Carpena02,Yamada04,Albuquerque05,Roche04} The
$1/k^{\alpha}$-law has also its trace in energy level
statistics.~\cite{Carpena04} All said above unambiguously
testifies that studying the properties of disordered systems with
long-range-correlated disorder is an attention grabbing task.

It is known since seminal papers by Bloch~\cite{Bloch28} and
Zener~\cite{Zener34} that an electron moving in a periodic potential
and subjected to a uniform electric field is localized due to the
Bragg reflection. It performs a periodic motion, known as Bloch
oscillations,~\cite{esa,Ashcr} which is characterized by an angular
frequency $\omega_{B}=eFd/\hbar$ and a spatial extension
$L_{B}=W/(eF)$, where $-e$ is the electron charge, $F$ is the applied
electric field strength, $d$ denotes the spatial period of the
potential, and $W$ stands for the band width. The Bloch oscillations
were observed for the first time as oscillations of electronic
wave-packets in semiconductor
superlattices~\cite{feld,leo,wasch,deko1,martini,los} (see for an
overview Ref.~\onlinecite{Leo98}), and later on as a periodic motion
of ensembles of ultracold atoms~\cite{dah,wilk} and Bose-Einstein
condensates~\cite{anders} in tilted optical lattices.  
The multiplicity  of observable physical phenomena related to
electron Bloch oscillations increases even more when a semiconductor
superlattice is subjected to joint, perpendicular or
tilted, electric and magnetic
fields.~\cite{kos1,kos2,bauer,Hummel05,kos3}

Recently, it was demonstrated that 1D disordered systems with the
$1/k^{\alpha}$ spectral density support Bloch-like oscillations of
quasiparticles.~\cite{Adame03} It was  also shown that these
oscillations provide a tool to measure the energy width of the
delocalized phase arising at $\alpha > \alpha_c$. More
specifically, the two mobility edges, which separate the phase of
extended states from the two phases of localized ones, were found
to play the role of effective band edges, i.e., it is the width
$\Delta$ of the mobility band who determines the spatial
extension $L^*_{B}= \Delta/(eF)$ of Bloch oscillations. Therefore,
this width can be measured in biased disordered lattices.

In this work we report further progress along this line by
considering the frequency-domain counterpart of the Bloch
oscillations, the so-called Wannier-Stark ladder
(WSL)~\cite{Wannier60} (see Refs.~\onlinecite{Mendez93} and
~\onlinecite{Chang93} for brief historical surveys). The WSL is
characterized by a series of equidistant quasistationary  levels
separated by an energy $U=\hbar\omega_{B}=eFd$. The progress in
semiconductor growth techniques made it possible to firmly
establish the existence of ladder level structures in
semiconductor
superlattices~\cite{Mendez88,Agullo89,Dignam90,Saker91} as well as
in $\delta$-doped superlattices.~\cite{Mendez94,Adame94} There
exists evidence that moderate uncorrelated disorder does not
destroy the required phase coherence to see the WSL in continuous
(many band) models.~\cite{Jose85} We focus on the one-band
approximation and demonstrate that in our system the WSL also
arises, in spite of the underlying randomness. Strong correlations
in the disorder facilitate the observation of the WSL. At $\alpha
> \alpha_c$, when the phase of extended states emerges at the
center of the band, the WSL may present a comb of doublets,
reflecting the doublet energy structure of the unbiased
system.~\cite{Diaz05} To work out the problem, we numerically
calculate the absorption spectrum varying the correlation exponent
$\alpha$, the disorder strength, and the magnitude $U$ of the
bias.

The outline of the paper is as follows. In the next section, we
present our model which is based on a tight-binding Hamiltonian of
a quasiparticle, moving in a long-range-correlated potential
landscape and subjected to a linear bias. In Sec.~\ref{absorption}
we recall the basic physics of the linear absorption spectrum in
the absence of both bias and correlations in disorder. The central
parts of the paper are Sec.~\ref{zeroU} and Sec.~\ref{nonzeroU},
where the results of numerical simulations of the absorption
spectrum profile in disorder-correlated systems are shown. We
begin with a brief discussion of the absorption spectrum behavior
upon increasing the disorder correlations in bias-free systems
(Sec.~\ref{zeroU}), proceeding in Sec.~\ref{nonzeroU} onto the
absorption in biased lattices. We discuss in detail its dependence
on the driving parameters of the model (bias magnitude, disorder
strength and correlation exponent $\alpha$) and provide an
evidence of that the correlations in disorder facilitate the
occurrence of the WSL. Finally, Sec.~\ref{conclusions} concludes
the paper.

\section{Model}
\label{model}

We consider a biased tight-binding model with diagonal disorder
on an otherwise regular 1D open lattice of spacing unity and $N$
sites ($N$ is assumed to be even). We assign to each lattice site
two levels (ground and  excited states) with transition energy
${\mathcal E}_n$ and consider optical transitions between them.
The model Hamiltonian is
\begin{eqnarray}
    {\cal H} & = & \sum_{n=1}^{N} \left[{\mathcal E}_n
    - U\left(n - \frac{N}{2}\right) \right]\,|n\rangle\langle n|
\nonumber\\
    & - & \sum_{n=1}^{N-1}\Big(|n\rangle\langle n+1|
    + |n+1\rangle\langle n|\Big)\ .
\label{hamiltonian}
\end{eqnarray}
Here, $|n\rangle$ denotes the state in which the $n$-th site is
excited, whereas all the other sites are in the ground state. The
energy of the state $|n \rangle$, ${\mathcal E}_n = \bar{\mathcal
E} + \varepsilon_n$, is assumed to have a stochastic part
$\varepsilon_n$ generated according to~\cite{Moura98}
\begin{equation}
    \varepsilon_{n} =\sigma C_\alpha \sum_{k=1}^{N/2}
    \frac{1}{k^{\alpha/2}}\, \cos\left(\frac{2\pi kn}{N}
    + \phi_k\right) \ ,
    \label{disorder}
\end{equation}
where $C_\alpha = \sqrt{2}\big(\sum_{k=1}^{N/2}
k^{-\alpha}\big)^{-1/2}$  is the normalization constant and
$\phi_1,\ldots,\phi_{N/2}$ are $N/2$ uncorrelated random phases
uniformly distributed within the interval $[0,2\pi]$. The
distribution~(\ref{disorder}) has zero mean $\langle
\varepsilon_{n}\rangle = 0$ and standard deviation $\langle
\varepsilon_{n}^2\rangle^{1/2} = \sigma$, where $\langle \ldots
\rangle$ indicates averaging over realizations of random phases
$\phi_k$. The quantity $\sigma$ will be referred to as magnitude
of disorder. The stochastic sequence~(\ref{disorder}) is
characterized by a correlation function
\begin{equation}
    \langle \varepsilon_{n} \varepsilon_m \rangle
    =  \frac{\sigma^2 C_\alpha^2}{2} \sum_{k=1}^{N/2}
    \frac{1}{k^\alpha}\,
    \cos\left[\frac{2\pi k(n-m)}{N}\right]\ .
\label{correlator}
\end{equation}
which is long-ranged, except for the particular value of the
exponent $\alpha = 0$, when $\langle \varepsilon_{n} \varepsilon_m
\rangle = \sigma^2 \delta_{mn}$, \, $\delta_{mn}$ being the
Kronecker $\delta$. The site energies in this case are
uncorrelated. Correlations, however, arise as soon as $\alpha \ne
0$. Thus, for $\alpha < 1$, they are power-law-like, i.e., the
correlation function~(\ref{correlator}) decays as $|n-m|^{\alpha
-1}$. If $\alpha \gg 1$, the term with $k = 1$ in the
series~(\ref{correlator}) is dominant, and $\langle
\varepsilon_{n} \varepsilon_m \rangle = (1/2)\sigma^2 C_\alpha^2
\cos [ 2\pi (n-m)/N ]$, implying full correlation of the on-site
energies.

The term $-U(n - N/2)$ in the Hamiltonian~(\ref{hamiltonian})
describes the linear bias. We will not necessarily relate it to
the presence of an external uniform electric field, as in the case
of an electron moving in the conduction band.~\cite{Dunlap86} The
bias can also be an intrinsic property of the system, as it takes
place in dendritic species (see, e.g., Ref.~\onlinecite{Heijs04}
and references therein). In this case, the
Hamiltonian~(\ref{hamiltonian}) models a 1D Frenkel exciton in a
disordered lattice with energetic bias. Finally, the intersite
transfer integrals in~(\ref{hamiltonian}) are restricted to
nearest-neighbors, and it is set to $-1$ over the entire lattice.
Also, we set $\bar{\mathcal E} = 0$ hereafter without loss of
generality.

As we already mentioned in the Introduction, in the absence of
bias $(U = 0)$, the above model supports a phase of extended
states at the center of the band, provided the correlation
exponent $\alpha$ is larger than a critical value $\alpha_c$. At
$\alpha < \alpha_c$ all the states are localized, which implies
that the model under consideration undergoes an LDT with respect
to the correlation exponent $\alpha$. In
Ref.~\onlinecite{Moura98}, where the above model of disorder was
introduced, the disorder magnitude was set to $\sigma = 1$, and
the critical value $\alpha_c$ was found to be $\alpha_c = 2$. It
may seem that $\alpha_c$ depends on $\sigma$. However, it was
demonstrated later on that $\alpha_c = 2$ is the universal
critical value for the LDT to occur in this model, independently
of $\sigma$.~\cite{Shima04}

Another peculiarity of this model, having a direct relationship
to the specific form of the random potential~(\ref{disorder}) as
well as to its localization properties, is that the absorption
spectrum at $\alpha > \alpha_c = 2$, i.e., in the presence of the
phase of extended states, reveals a double-peaked structure (see
Ref.~\onlinecite{Diaz05} and Sec.~\ref{zeroU}).

\section{Absorption spectrum}
\label{absorption}

The quantity subjected to calculation throughout this paper will
be the absorption spectrum defined as
\begin{equation}
    A(E)= \frac{1}{N}\Bigg\langle \sum_{\nu=1}^{N}
    \left( \sum_{n=1}^{N} \psi_{\nu n} \right)^{2}
    \delta(E - E_{\nu})\Bigg\rangle \ ,
\label{lineshape}
\end{equation}
where the $E_\nu$ and $\psi_{\nu n}$ are the eigenenergies and
eigenfunctions, respectively, obtained after diagonalization of
the Hamiltonian~(\ref{hamiltonian}). The quantity $F_\nu = \left(
\sum_{n=1}^{N} \psi_{\nu n} \right)^{2}$ is the dimensionless
oscillator strength of the $\nu$-th state.

In order to gain insight into the effects of long-range
correlations in disorder and the bias on the absorption, we first
recall the basic features of the absorption spectrum in the
absence of both ($\alpha = 0$ and $U = 0$). Uncorrelated diagonal
disorder results in localization of all the
states~\cite{Abrahams79} and in the appearance of Lifshits tails
in the density of states (DOS), outside the bare quasi-particle
band which ranges from $E = -2$ to $E = 2$. Since we set negative
sign of the hopping integral, the low-energy part of the DOS (with
$E$ around $-2$) is of importance for the linear optical
absorption. The majority of states lying in this region are
localized at different (weakly overlapped) segments. Some of them
are bell-like, i.e., without nodes within localization region,
while the other and higher (band) states resemble standing waves
with nodes.~\cite{Malyshev95} The bell-like states dominate the
optical absorption because they accumulate oscillator strengths
large compared to those of the other states. They result from
localization on fluctuations of the site potential, which have a
well-like envelop.~\cite{Malyshev:unpublished} The typical size of
a such potential wells, $N^*$, determines the extension of the
bell-like states, while the depth $\sigma/\sqrt{N^*}$ governs the
width of the absorption spectrum (see below). Notice that the
potential depth is $\sqrt{N^*}$ times as small as the bare
magnitude $\sigma$. This effect is known as the exchange (or
motional) narrowing: fluctuations of a stochastic site potential of
alternating signs are averaged out by a quasiparticle rapidly
moving (due to a large exchange interaction $J \gg \sigma$) within a
region of size
$N^*$.~\cite{Malyshev95,Fidder91,Malyshev91,Knapp84,Malyshev99}

The absorption spectrum is peaked slightly below the bare band
edge $E = -2$ and represents an inhomogeneously broadened line
with a Gaussian-like red and a Lorentzian-like blue tail (see,
e.g., Ref.~\onlinecite{Fidder91}). The localization size of the
optically dominant states, $N^*$, and the full width at half
maximum (FWHM) of the absorption, $\sigma^*$, are estimated
as~\cite{Malyshev91}
\begin{subequations}
\label{N*sigma*}
\begin{equation}
\label{N*}
    N^* = \left( \frac{3\pi^2}{\sigma}\right)^{2/3} \ ,
\end{equation}
\begin{equation}
\label{sigma*}
    \sigma^* = \frac{2\sigma}{\sqrt{N^*}}
    = 6\pi^2 \left(\frac{\sigma}{3\pi^2}\right)^{4/3} \ .
\end{equation}
\end{subequations}

To conclude this section, note that both short-range and power-law
long-range correlations in disorder result in decreasing the
localization length of the band edge
states~\cite{Russ98,Malyshev99} and, subsequently, increasing the
absorption line width.~\cite{Diaz05,Malyshev99} In contrast, this
is not the case for correlations which are stronger than
power-law-like. An example is a sequence~(\ref{disorder}) at
$\alpha \gg 1$ (see the next section).

\section{Unbiased system} \label{zeroU}

\begin{figure}
\centering
\includegraphics[width=70mm,clip]{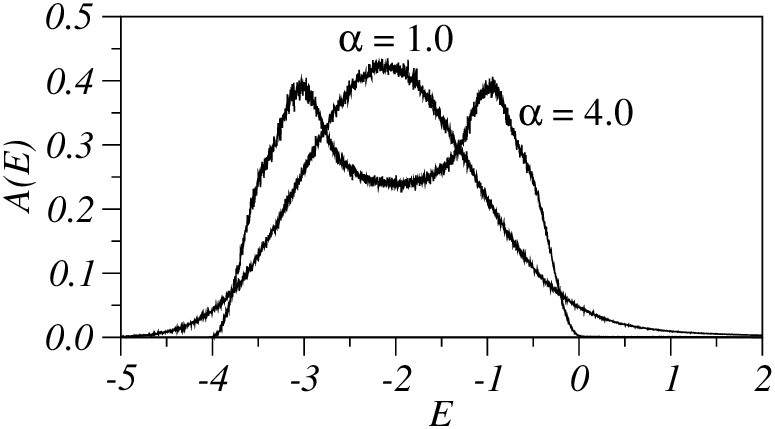}
\caption{Absorption spectra of an ensemble of unbiased chains ($U
= 0$) with $N = 500$ sites calculated for two values of the
correlation exponent $\alpha$, shown in the plot. In both cases,
the magnitude of disorder is $\sigma = 1$. Each curve were
obtained after averaging over $3 \times 10^4$ realizations of
disorder.}
\label{fig1}
\end{figure}

We first discuss the absorption spectrum of the unbiased system
($U =0$), aiming to separate the effects of bias from those
related to the disordered nature of the model. Figure~\ref{fig1}
represents the absorption spectra calculated for two values of the
correlation exponent, $\alpha = 1$ and $\alpha = 4$. By
convention, we will refer to these two cases as weakly and
strongly long-range correlated disorder, respectively. The results
were obtained by numerically diagonalizing the
Hamiltonian~(\ref{hamiltonian}) for chains of size $N = 500$ with
open boundary conditions. The disorder magnitude was set to
$\sigma = 1$. Averaging over $3 \times 10^{4}$ realizations of the
disorder were applied in Eq.~(\ref{lineshape}) for each value of
$\alpha$.

From Fig.~\ref{fig1} we observe that in the case of weak
correlations in disorder ($\alpha = 1$), the absorption spectrum
consists of a single inhomogeneously broadened asymmetric line.
Its red and blue tails can be fitted by a Gaussian and Loretzian,
respectively. The spectrum is peaked slightly below the low-energy
band edge $E = -2$ and has the FWHM $\sigma^* \approx 2.2$ which
is approximately three times larger than that in the limit of
uncorrelated disorder ($\alpha = 0$), Eq.~(\ref{sigma*}). This
tendency is in full accordance with that we mentioned in the
preceding section.

The absorption line shape, however, changes dramatically when
disorder is strongly long-range correlated ($\alpha = 4$). The
higher-energy peak in the absorption spectrum starts to build up
when the correlation exponent $\alpha$ exceeds the critical value
$\alpha_c = 2$ or, in other words, when a phase of the extended
states emerges in the center of the band.~\cite{Diaz05} Its
appearance implies that states deep inside the band gain large
oscillator strengths. This is in contrast to the case of
uncorrelated disorder, where higher states have a vanishingly
small oscillator strength.~\cite{Fidder91,Malyshev91}

A simple explanation of this anomaly is based on the fact that
for sufficiently large values of $\alpha$, the first term in the
series~(\ref{disorder}) is dominant, while the others are
considerably smaller. Consequently, the site potential for a given
realization is cosine-like (harmonic with $k=1$), perturbed by a
colored noise (harmonics with $k \geq 2$). Then, the whole lattice
can be represented as a two weakly-coupled sublattices with
different site energy (see Ref.~\onlinecite{Diaz05} for more
details). Optical transitions to the band edge states of these two
sublattices give rise to a double-peaked structure of the
absorption spectrum. Remarkably, the higher-energy peak monitors
the upper mobility edge of the delocalized phase.~\cite{Diaz05}

\section{Biased system} \label{nonzeroU}

\subsection{Qualitative picture}
\label{qualitative}

In disorder-free systems, switching the bias on results in
dynamical (Bloch) localization of all the states~\cite{Bloch28}
within the localization size $L_B = 4/U$, where 4 is the band
width in dimensionless units. The Bloch localization is
accompanied by the subsequent reorganization of the energy
spectrum of the system, which becomes ladder-like with the level
spacing $U$.~\cite{Wannier60} This structure is revealed in
photoluminescence~\cite{Mendez88,Agullo89} and
photoconductivity~\cite{Saker91} spectra as a series of equally
spaced peaks. Disorder broadens the peaks and makes them
unresolved. Below we derive a relationship between magnitudes of
disorder $\sigma$ and bias $U$, which governs the occurrence of
the WSL in disordered systems.

Briefly, our reasoning is as follows (a detailed study will be
published elsewhere~\cite{Malyshev06}). According to the exchange
narrowing concept (see the discussion in Sec.~\ref{absorption} and
Refs.~\onlinecite{Malyshev95,Fidder91,Malyshev91,Knapp84,Malyshev99}),
a quasiparticle confined within a chain segment of size $L_B$ sees
a disorder of a reduced magnitude $\sigma/\sqrt{L_B}$, with
$\sigma$ the strength of the bare disorder. Therefore, the
resulting inhomogeneous broadening of the WSL levels can be
estimated, similarly to Eq.~(\ref{sigma*}), as
\begin{equation}
\label{WSLsigmaB}
    \sigma_B = \frac{2\sigma}{\sqrt{L_B}} = \sigma \sqrt{U} \ .
\end{equation}
In order to resolve the ladder structure, the inhomogeneous width
$\sigma_B$ must be smaller than the level spacing $U$. This brings
us to the condition,
\begin{equation}
\label{WSLoccur}
    \sigma < \sqrt{U} \ ,
\end{equation}
which governs the occurrence of the WSL in 1D disordered systems. We
will refer to the relationship~(\ref{WSLoccur}) as to the limit of
strong bias.

Now, we turn to discussing the opposite sign of
inequality~(\ref{WSLoccur}), which we will name the low or/and
moderate bias limit. At non zero bias, each site gets an
additional energy $-U(n -N/2)$. Thus, the energy difference
between the edge sites (magnitude of the total bias) is about
$UN$. It is to be compared first of all to the absorption line
width $\sigma^*$ in the absence of bias. Apparently, at $UN \ll
\sigma^*$, i.e., when the potential drop across the whole system
is much smaller than the width, the effect of bias on the
absorption is negligible. On the contrary, the bias is expected to
broaden the absorption spectrum when $UN \gg \sigma^*$. Indeed,
optically dominant (bell-like) states in this case will be
distributed from $[-UN/2$ to $UN/2]$, giving rise to almost
constant absorption within this energy range.

Such a scenario, however, holds as long as $UN^* < \sigma^*$,
i.e.,  provided the typical potential drop across the potential
wells supporting bell-like states is smaller than the typical well
depth $\sigma^*$ in bias-free systems. At $UN^* > \sigma^*$, the
picture of localization in the potential wells does not work
anymore. The localization of states now is governed by a
complicated interplay of disorder and bias, which is hard to
handle qualitatively. Nevertheless, we can still claim that the
FWHM of the absorption spectrum will be on the order of magnitude
of the total bias, $UN$. On further increasing the bias, we fall
in the regime of the WSL, $\sqrt{U} > \sigma$. Our numerical
simulation confirm this qualitative picture, as shown below.

Note that the above reasonings can be applied without any remarks
to both uncorrelated and weakly correlated disorder ($\alpha <
1$). In the limit of strong correlations ($\alpha \gg 1$) the
picture requires corrections which we discuss in
Sec.~\ref{largeU}.

\begin{figure}[ht]
\centerline{\includegraphics[width=70mm,clip]{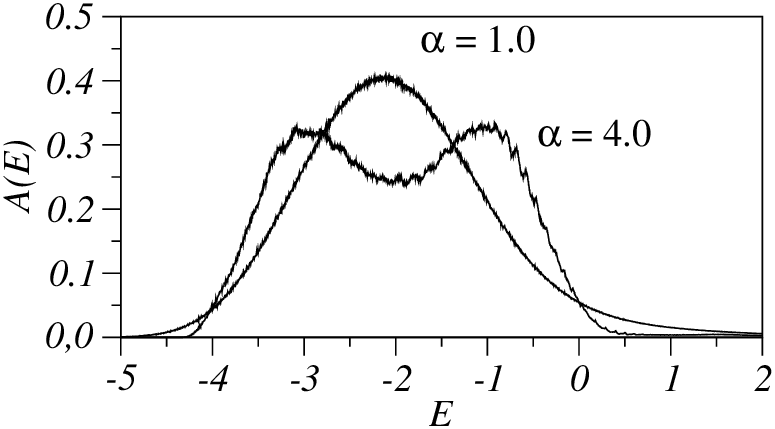}}
\caption{Absorption spectra of biased chains ($U = 0.01$) with  $N
= 100$ sites calculated for two values of the correlation exponent
$\alpha$, shown in the plot. The magnitude of disorder is $\sigma
= 1$. Each curve were obtained after averaging over $10^6$
realizations of disorder.} \label{fig2}
\end{figure}

\subsection{Low and moderate bias}
\label{smallU}

In Fig.~\ref{fig2} we plotted the absorption spectra calculated
for disordered biased chains of $N = 100$ sites, choosing the
disorder strength $\sigma = 1$ and the bias magnitude $U = 0.01$.
Averaging over $10^6$ realizations of disorder were performed in
Eq.~(\ref{lineshape}). Two values of the correlation exponents
were considered, $\alpha = 1$ and $\alpha = 4$. As $\sigma \gg
\sqrt{U}$, we are not in the WSL regime. Furthermore, at $U =
0.01$ the overall potential drop $UN = 1$ is smaller than the FWHM
in the absence of bias, $\sigma^* \approx 2.2$ (see
Fig.~\ref{fig1}). As a consequence, the effect of bias is weak,
leading only to a unnoticeable broadening of the spectra and
smoothing the shape of the doublet (at $\alpha = 4$) as compared
to the bias-free conditions (compare to Fig.~\ref{fig1}).

\begin{figure}[ht]
\centerline{\includegraphics[width=70mm,clip]{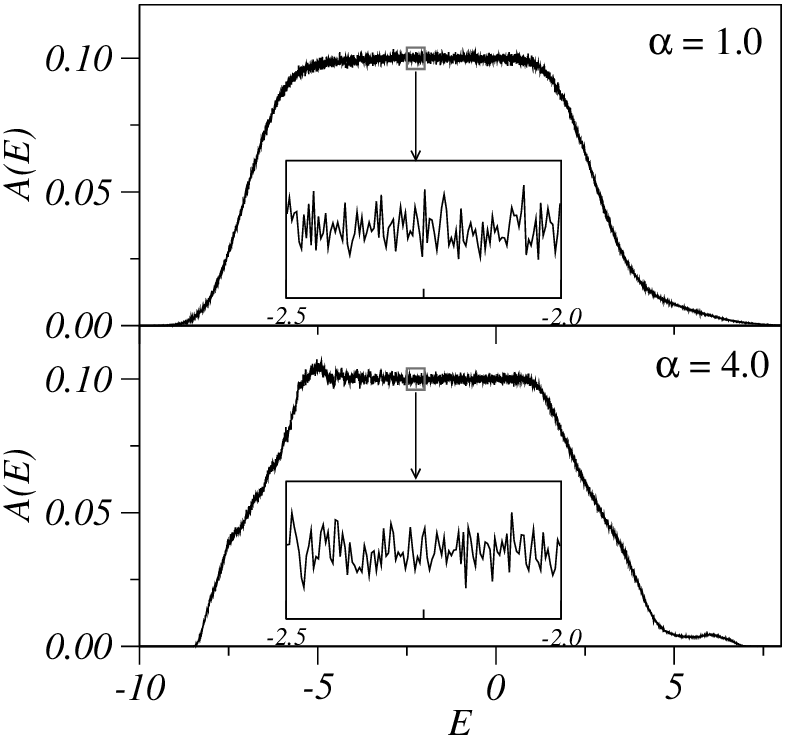}}
\caption{ Absorption spectra of biased chains ($U = 0.1$) with $N
= 100$ sites calculated for two values of the correlation exponent
$\alpha$, shown in the plot. The magnitude of disorder is $\sigma
= 1$. Each curve were obtained after averaging over $10^6$
realizations of disorder. Insets show enlarged views of the
spectra within the grey boxes.} \label{fig3}
\end{figure}

Figure~\ref{fig3} presents the results of the simulations for a
larger magnitude of the bias, $U = 0.1$, keeping all other
parameters unchanged. Still, $\sigma > \sqrt{U}$ that is not in
favor of the WSL. However, the total bias $UN = 10$ is now larger
than the bias-free FWHM $\approx 2.2$. Therefore, for both values
of the correlation exponent $\alpha =1$ and $\alpha = 4$ the
absorption spectrum shows a plateau-like shape and a large
broadening, with the FWHM $\approx 10$ equal to the total bias.
These trends are in full agreement with our qualitative reasoning
presented in the preceding section.

\begin{figure}[ht]
\centerline{\includegraphics[width=70mm,clip]{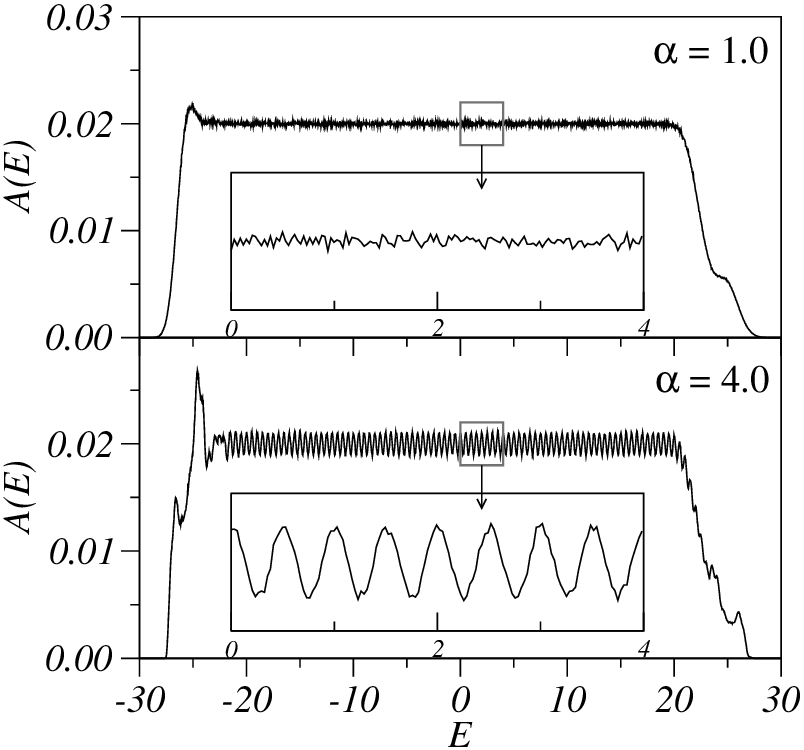}}
\caption{Same as in Fig.~\ref{fig3}, but for a bias $U = 0.5$.}
\label{fig4}
\end{figure}

\begin{figure}[ht]
\centerline{\includegraphics[width=70mm,clip]{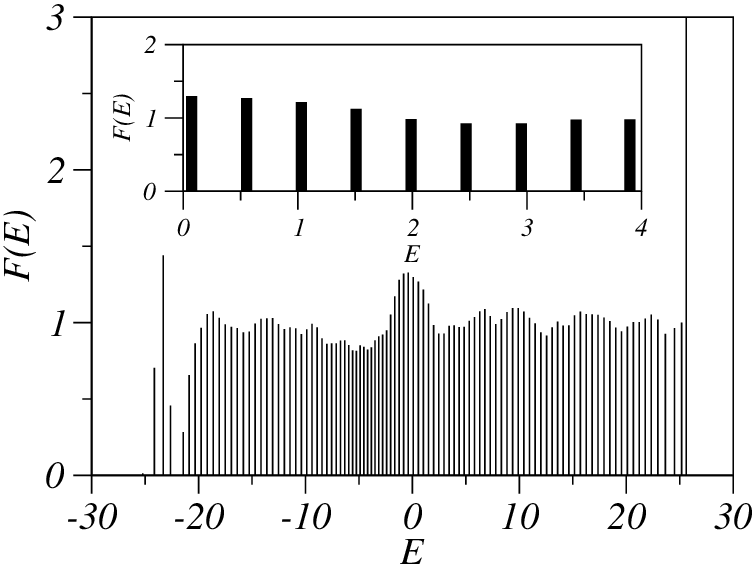}}
\caption{The oscillator strength distribution for the set of
parameters as in lower panel of Fig.~\ref{fig4}.} \label{fig5}
\end{figure}

\subsection{Strong bias}
\label{largeU}

Aiming to find fingerprints of the WSL in the optical absorption
spectra, we further increased the magnitude of the bias.
Figure~\ref{fig4} shows the spectra calculated for the bias
magnitude $U = 0.5$. As before, chains of $N = 100$ sites were
used in the simulations and two values of the correlation exponent
$\alpha$ were considered, $\alpha = 1$ and $\alpha = 4$. The
disorder strength was set to $\sigma = 1$, and averaging over
$10^6$ realization of disorder was performed in
Eq.~(\ref{lineshape}).

\begin{figure}[ht]
\centerline{\includegraphics[width=70mm,clip]{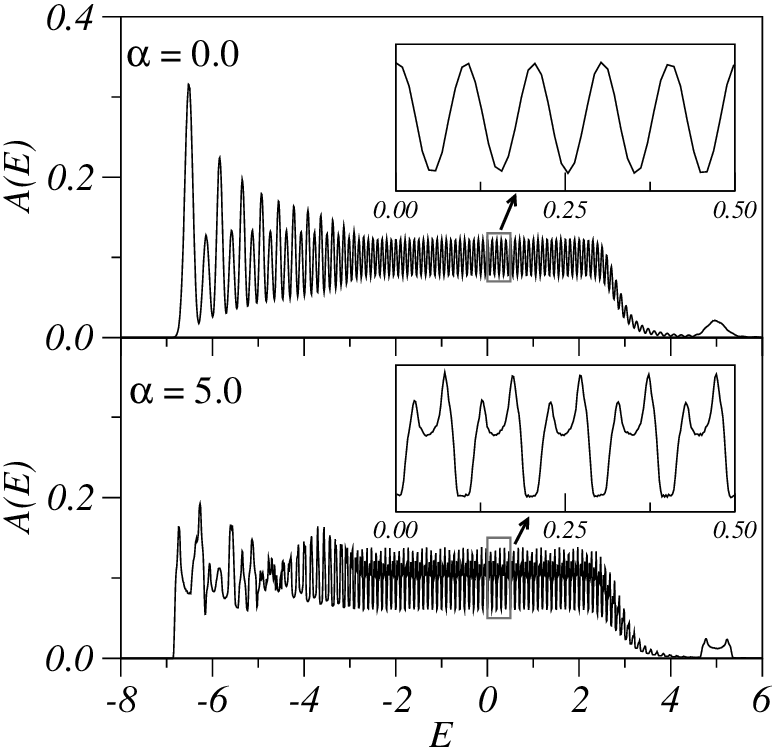}}
\caption{Same as in Fig.~\ref{fig3}, but for a disorder magnitude
$\sigma = 0.2$. The inset shows an enlarged view of the center of
the band.} \label{fig6}
\end{figure}

We observe (the upper panel) that at $\alpha =1 < \alpha_c = 2$
the spectrum remains structureless that is consistent with the
estimate~(\ref{WSLoccur}): still $\sigma > \sqrt{U}$, i.e., the
WSL can not be resolved. However, for strong correlations in
disorder, when $\alpha = 4 > \alpha_c = 2$, the spectrum presents
a periodic pattern which is not masked by the stochastic disorder
fluctuations (see the inset in the lower panel). Most important,
the period of the modulation is exactly equal to $U = 0.5$.  To
illustrate more our statement, we depicted in Fig.~\ref{fig5} the
oscillator strength $F_{\nu}$ as a function of the eigenenergy
$E_{\nu}$ for $\alpha = 4$. It is clearly seen the periodicity in
energy with a period equal to $U = 0.5$ as well as the similarity
of the corresponding oscillator strengths. We checked out that the
pattern with the same spacing also holds for larger systems, i.e,
its period is not a consequence of finite size effects. Therefore,
we claim that the periodic pattern found in the simulations
results from the occurrence of the WSL in the energy spectrum of
the system.

An explanation of the occurrence of the WSL in the strong
correlation regime ($\alpha = 4$) is based on the already
mentioned in Sec.~\ref{zeroU} fact: the site
potential~(\ref{disorder}) for a given realization of disorder is
cosine-like (harmonic with $k=1$), perturbed by a colored noise
(harmonics with $k \geq 2$). This implies that the magnitude of
the actual disorder is at least by a factor of $2^{-\alpha/2}$
smaller than the bare value $\sigma$; the
inequality~(\ref{WSLoccur}) turns out to be fulfilled for the
reduced disorder. In other words, strong correlations in disorder
facilitate the occurrence of the WSL.

Finally, we calculated the absorption spectrum profile for
magnitudes of disorder $\sigma = 0.2$ and bias $U = 0.1$, when the
inequality~(\ref{WSLoccur}) holds even in the absence of
correlations in disorder ($\alpha = 0$). The results, obtained for
$\alpha = 0$ and $\alpha = 5$ are depicted in Fig.~\ref{fig6}
(upper and lower panels, respectively). One observes that the
absorption spectrum shows a resolved structure for both values of
$\alpha$, which, however, is heterogeneous. The central part of
the spectrum is of our primary interest because it reveals a
periodic pattern.

At $\alpha = 0$ (uncorrelated disorder), the pattern consists of
equally spaced single peaks with the spacing exactly equal to $U =
0.1$ (see the inset). This again allows us to associate the peaks
with the occurrence of WSL in the energy spectrum of the system.
The FWHM of a single peak is about 0.06 that is in good agreement
with Eq.~(\ref{WSLsigmaB}). We checked also out that the pattern
did not appear if $\sigma = 0.4 \approx \sqrt{U} = 0.45$, in full
accordance with our reasonings presented in
Sec.~\ref{qualitative}. These results corroborate those found in
Ref.~\onlinecite{Jose85} for random Kronig-Penney models.

In the case of strongly correlated disorder ($\alpha = 5$), each
single peak of the central pattern splits into a doublet, as is
seen from the inset in the lower panel. The origin of the doublets
can be traced back to the behavior of the unbiased system, where
the absorption line shape exhibits a doublet structure provided
$\alpha > \alpha_c = 2$ (see Sec.~\ref{zeroU} and
Ref.~\onlinecite{Diaz05}). Figure~\ref{fig6} points out that the
splitting also occurs in the presence of bias.

To conclude we comment on the peculiarities of the red and blue sides
of the absorption spectrum. We associate them  with finite size
efects. Indeed, these parts of the energy spectrum are formed by the
states localized close to the system ends. Because of that, the
corresponding eigenfunctions differs from those at the band center.

\section{Summary and concluding remarks}

\label{conclusions}

We studied numerically the linear optical absorption of a
quasiparticle moving on a 1D disordered lattice subjected to a
linear bias of magnitude $U$. The random site potential was set to
have a power-law spectral density $S(k)\sim 1/k^{\alpha}$, which
gives rise to long-range correlations in site energies.

The absorption spectrum of the unbiased lattice ($U = 0$) was
found to present a single peak in the weakly-correlated limit,
when the correlation exponent $\alpha < \alpha_c = 2$, with
$\alpha_c = 2$ the critical value for the occurrence of a phase of
extended states in the center of the band. For strongly-correlated
disorder, at $\alpha > \alpha_c = 2$, the absorption lineshape
turned out to exhibit a double-peaked structure, characteristic
for this type of long-range correlations.~\cite{Diaz05}

Switching the bias on does not change much the outlined behavior
provided the magnitude of overall bias $UN$ does not exceed the
absorption bandwidth of the unbiased system. At higher magnitudes
of bias, the absorption spectrum starts to broaden, independently
of the magnitude of the correlation exponent $\alpha$. Its profile
gets a top hat shape, being almost flat at the center of the
absorption band and having the FWHM on the order of $UN$. Such a
scenario holds as long as the disorder magnitude $\sigma <
\sqrt{U}$.

On further increasing the magnitude of the bias, a periodic
pattern is found to build up at the center of the (already wide)
absorption band. Its period is equal to $U$, as for the
Wanier-Stark ladder in an ideal lattice, and independent of the
system size $N$. Therefore, we attribute the pattern found to the
Wannier-Stark quantization of the energy spectrum in the
disordered lattice.

The occurrence of the Wannier-Stark pattern is facilitated by the
presence of correlations in disorder. In the limit of strong
correlations ($\alpha > \alpha_c = 2$), each Wannier-Stark level
represents a doublet, reflecting the doublet structure of the
absorption spectrum of the unbiased system.

To conclude we note that in our study we did not explicitly relate
the bias to the presence of an external uniform electric field.
Therefore, our conclusions can be equally applied both to 1D
disordered electrons, moving in a uniform electric field, and to
1D disordered Frenkel excitons, where energetic bias can be an
intrinsic property of the system. Dendrimers represent one of the
examples.~\cite{Heijs04}

\acknowledgments

Work at Madrid was supported by 
MEC (Project MAT2003-01533). Yu.A.\ K. acknowledges
support from Generalitat Valenciana (Grant CTESIN/2005/022). V.\
A.\ M.\ acknowledges support from NanoNed, a national
nanotechnology programme coordinated by the Dutch Ministry of
Economic Affairs, and from ISTC (grant \#2679).


\begin{thebibliography}{99}

\bibitem{Moura98} F.\ A.\ B.\ F.\ de Moura and M.\ L.\ Lyra, Phys.\ Rev.\
    Lett.\ \textbf{81}, 3735 (1998); Physica A, \textbf{266}, 465 (1999).

\bibitem{Izrailev99} F.\ M.\ Izrailev and A.\ A.\ Krokhin, Phys.\ Rev.\
    Lett.\ \textbf{82}, 4062 (1999).

\bibitem{Liu99} W.-S.\ Liu, T.\ Chen, and S.-J.\ Xiong, J.\ Phys.\ Condens.\
    Matter \textbf{11}, 6883 (1999).

\bibitem{Kantelhardt00}J.\ W.\ Kantelhardt, S.\ Russ, A.\ Bunde, S.\ Havlin,
    and I.\ Webman, Phys. Rev. Lett. \textbf{84}, 198 (2000); {\it ibid.}
    F.\ A.\ B.\ F.\ de Moura and M.\ L.\ Lyra, \textbf{84}, 199 (2000).

\bibitem{Kuhl00} U.\ Kuhl, F.\ M.\ Izrailev, A.\ A.\ Krokhin, and H.\ -J.\
    St\"{o}ckmann, Appl.\ Phys.\ Lett.\ \textbf{77}, 633 (2000).

\bibitem{Russ01} S.\ Russ, J.\ W.\ Kantelhardt, A.\ Bunde, and S.\ Havlin,
    Phys. Rev. B \textbf{64}, 134209 (2001).

\bibitem{Lindquist01} B.\ Lindquist, Phys.\ Rev.\ E \textbf{63}, 056605 (2001).

\bibitem{Chen02} X.\ Chen and T.\ Kobayashi, Solid State Commun.\ \textbf{122},
     479 (2002).

\bibitem{Zhang02} G.-P.\ Zhang and S.-J.\ Xiong, Eur.\ Phys.\ J.\ B
    \textbf{29}, 491 (2002).

\bibitem{Carpena02}P.\ Carpena, P.\ Bernaola-Galv\'{a}n, P.\ Ch.\ Ivanov, and
    H.\ E.\ Stanley, Nature {\bf 418}, 955 (2002); {\it ibid} {\bf 421}, 764
    (2003).

\bibitem{Lima02} R.\ P.\ A.\ Lima, M.\ L.\ Lyra, E.\ M.\ Nascimento, and A.\ D.\
    de Jesus, Phys.\ Rev.\ B \textbf{65}, 104416 (2002).

\bibitem{Moura02} F.\ A.\ B.\ F.\ de Moura, M.\ D.\ Coutinho-Filho, E.\ R.\
    Raposo, and M.\ L.\ Lyra, Phys.\ Rev.\ B \textbf{66}, 014418 (2002);
    {\it ibid.} \textbf{68}, 012202 (2003).

\bibitem{Adame03} F.\ Dom\'{\i}nguez-Adame, V.\ A.\ Malyshev, F.\ A.\ B.\ F.\
    de Moura, and M.\ L.\ Lyra, Phys.\ Rev.\ Lett.\ \textbf{91}, 197402
    (2003).

\bibitem{Liu03} W.-S.\ Liu, S.\ Y.\ Liu, and X.\ L.\ Lei, Eur.\ Phys.\ J.\ B
    \textbf{33}, 293 (2003).

\bibitem{Carpena04} P.\ Carpena, P.\ Bernaola-Galv\'{a}n, and P.\ Ch.\
    Ivanov, Phys.\ Rev.\ Lett.\ \textbf{93}, 176804 (2004).

\bibitem{Moura04} F.\ A.\ B.\ F.\ de Moura, M.\ D.\ Coutinho-Filho, M.\ L.\
    Lyra, and E.\ P.\ Raposo, Europhys.\ Lett.\ \textbf{66}, 585
    (2004).

\bibitem{Ndawana04}M. L. Ndawana , R. A. R\"{o}mer, and M. Schreiber,
    Europhys. Lett. \textbf{68}, 678 (2004).

\bibitem{Yamada04} H.\ Yamada, Phys.\ Lett.\ A \textbf{332}, 65 (2004); Int.\
    J.\ Mod.\ Phys.\ B \textbf{18}, 1697 (2004);  Phys.\ Rev.\ B
    \textbf{69}, 014205 (2004).

\bibitem{Shima04} H.\ Shima, T.\ Nomura, and T.\ Nakayama, Phys.\ Rev.\ B
    \textbf{70}, 075116 (2004).

\bibitem{Albuquerque05} E.\ L.\ Albuquerque, M.\ S.\ Vasconcelos, M.\ L.\
    Lyra, and F.\ A.\ B.\ F.\ de Moura, Phys.\ Rev.\ E, \textbf{71},
    021910 (2005).

\bibitem{Cheraghchi05} H.\ Cheraghchi, S.\ M.\ Fazeli, and K.\ Esfarjani,
    Phys.\ Rev.\ B \textbf{72}, 174207 (2005).

\bibitem{Russ98}S. Russ, S. Havlin, and I. Webman, Phil. Mag. B \textbf{77},
    1449 (1998); S. Russ, J. W. Kantelhardt, A. Bunde, S. Havlin, and I.
    Webman, Physica A 266, 492 (1999).

\bibitem{Abrahams79} E.\ Abrahams, P.\ W.\ Anderson, D.\ C.\ Licciardello,
    and T.\ V.\ Ramakrishnan, Phys.\ Rev.\ Lett.\ {\bf 42}, 673 (1979).

\bibitem{Lee85} P.\ A.\ Lee and T.\ V.\ Ramakrishnan, Rev.\ Mod.\
    Phys.\ {\bf 57}, 287 (1985).

\bibitem{Kramer93} B.\ Kramer and A.\ MacKinnon, Rep.\ Prog.\ Phys.\
    {\bf 56}, 1469 (1993).

\bibitem{Beenakker97} C.\ W.\ J.\ Beenakker, Rev. Mod. Phys. {\bf
    69}, 731 (1997).

\bibitem{Janssen98} M.\ Janssen, Phys. Rep. {\bf 295}, 1 (1998).

\bibitem{Paczuski96} M.\ Paczuski, S. Maslov, and P.\ Bak, Phys.\
    Rev.\ E \textbf{53}, 414 (1996).

\bibitem{Havlin99} S.\ Havlin, S.\ V.\ Buldyrev, A.\ Bunde, A.\ L.\
    Goldberger, P.\ Ch.\ Ivanov, C.-K.\ Peng, and H.\ E.\ Stanley, Physica
    A \textbf{273}, 46 (1999).

\bibitem{Mandelbrot82} B.\ B.\ Mandelbrot, {\it The Fractal Geometry in
    Nature}, Freeman, San Francisco, 1982.

\bibitem{Peng92} C.-K.\ Peng, S.\ V.\ Buldyrev, A.\ L.\ Goldberger,
    S.\ Havlin, F.\ Sciortino, M.\ Simons, and H.\ E.\ Stanley,, Fractals
    \textbf{1}, 283 (1993).

\bibitem{Li92} W.\ Li and K.\ Kaneko, Europhys. Lett. \textbf{17}, 655 (1992).

\bibitem{Voss92} R.\ Voss, Phys. Rev. Lett. \textbf{68}, 3805 (1992).

\bibitem{Stanley93} H.\ E.\ Stanley, S.\ V.\ Buldyrev, A.\ L.\ Goldberger,
    S.\ Havlin, S.\ M. Ossadnik, C.-K.\ Peng, and M.\ Simons, Fractals
    \textbf{1}, 283 (1993).

\bibitem{Buldyrev98} S.\ V.\ Buldyrev, N.\ V.\ Dokholyan, A.\ L.\ Goldberger,
    S.\ Havlin,  C.-K.\ Peng, H.\ E.\ Stanley, and G.\ M.\ Viswanathan,
    Physica A \textbf{249}, 430 (1998).

\bibitem{Roche04} S.\ Roche, D.\ Bicout, and E.\ Maci\'{a}, Phys. Rev.\ Lett.\ B
   {\bf 91}, 109901 (2004).

\bibitem{Bloch28} F.\ Bloch, Z. Phys. \textbf{52}, 555 (1928).

\bibitem{Zener34} C.\ Zener, Proc.\ R.\ Soc.\ London, Ser.\ A \textbf{145}, 523
    (1934).

\bibitem{esa} L. Esaki and R. Tsu, IBM J. Res. Div. \textbf{14},
    61 (1970).

\bibitem{Ashcr} N. W. Ashcroft and N. D. Mermin,{\it Solid
    State Physics} (Saunders Colledge Publishers, New York, 1976),
    p. 213.

\bibitem{feld} J. Feldmann, K. Leo, J. Shah, D. A. B. Miller,
    J. E. Cunningham, T. Meier, G. von Plessen, A. Schulze,
    P. Thomas, and S. Schmitt-Rink, Phys. Rev. B \textbf{46},
    R7252 (1992).

\bibitem{leo} K. Leo, P. Haring Bolivar, F. Br\"{u}ggemann,
    R. Schwedler, and K. K\"{o}hler, Solid State Commun. \textbf{84},
    943 (1992).

\bibitem{wasch} C. Waschke, H. G. Roskos, R. Schwedler, K. Leo,
    H. Kurz, and K. K\"ohler, Phys. Rev. Lett. \textbf{70}, 3319
    (1993).

\bibitem{deko1} T. Dekorsy, P. Leisching, K. K\"{o}hler, and
    H. Kurz, Phys. Rev. B \textbf{50}, R8106 (1994).

\bibitem{martini} R. Martini, G. Klose, H. G. Roskos, H. Kurz,
    H. T. Grahn, and R. Hey, Phys. Rev. B \textbf{54}, R14325
    (1996).

\bibitem{los} F. L\"{o}ser, Yu. A. Kosevich, K. K\"{o}hler,
    and K. Leo, Phys. Rev. B \textbf{61}, R13373 (2000).

\bibitem{Leo98} K. Leo, Semicond. Sci. Technol. \textbf{13},
    249 (1998).

\bibitem{dah} M. BenDahan, E. Peik, J. Reichel, Y. Castin, and C.
    Salomon, Phys. Rev. Lett. \textbf{76}, 4508 (1996).

\bibitem{wilk} S. R. Wilkinson, C. F. Bharucha, K. W. Madison,
    Q. Niu, and M. G. Raizen, Phys. Rev. Lett. \textbf{76}, 4512 (1996).

\bibitem{anders} B. P. Anderson and M. A. Kasevich, Science
    \textbf{282}, 1686 (1998).

\bibitem{kos1} Yu. A. Kosevich, Phys. Rev. B \textbf{63},
    205313 (2001).

\bibitem{kos2} Yu. A. Kosevich, Phys. Rev. Lett. \textbf{88},
    229701 (2002).

\bibitem{bauer} T. Bauer, J. Kolb, A. B. Hummel, H. G. Roskos,
    Yu. Kosevich, and K. K\"ohler, Phys. Rev. Lett. \textbf{88},
    086801 (2002).

\bibitem{Hummel05} A. B. Hummel, C. Bl\"oser, T. Bauer, H. G.
    Roskos, Yu. A. Kosevich, and K. K\"ohler, Phys. Stat. Sol. B
    \textbf{242}, 1175 (2005).
    
\bibitem{kos3} Yu. A. Kosevich, A. B. Hummel, H. G. Roskos,
     and K. K\"ohler, Phys. Rev. Lett. \textbf{96}, 137403 (2006).    

\bibitem{Wannier60} G.\ H.\ Wannier, Phys. Rev. \textbf{117}, 432 (1960).

\bibitem{Mendez93} E.\ E.\ M\'endez and G.\ Bastard, Phys. Today \textbf{46}
    (6), 34 (1993).

\bibitem{Chang93} M.\ C.\ Chang and Q.\ Niu, Phys.\ Rev.\ B {\bf 48}, 2215
    (1993).

\bibitem{Mendez88} E.\ E.\ M\'endez, F.\ Agull\'{o}-Rueda, and J.\ M.\ Hong,
    Phys.\ Rev.\ Lett.\ {\bf 60}, 2426 (1988).

\bibitem{Agullo89} F.\ Agull\'{o}-Rueda, E.\ E.\ M\'endez, and J.\ M.\ Hong,
    Phys.\ Rev.\ B {\bf 40}, 1357 (1989).

\bibitem{Dignam90} M.\ M.\ Digman and J.\ E.\ Sipe, Phys.\ Rev.\ Lett.\ {\bf
    64}, 1797 (1990).

\bibitem{Saker91} M.\ K.\ Saker, D.\ M.\ Whitteker, M.\ S.\ Skolnick, M.\ T.\
    Emeny, and C.\ R.\ Whitehouse, Phys.\ Rev.\ B {\bf 43}, 4945 (1991).

\bibitem{Mendez94} B.\ M\'{e}ndez and F.\ Dom\'{\i}nguez-Adame, Phys.\ Rev.\ B
    \textbf{49}, 11\,471 (1994).

\bibitem{Adame94} F.\ Dom\'{\i}nguez-Adame, B.\ M\'{e}ndez, and E.\ Maci\'{a},
    Semicond.\ Sci.\ Technol.\ \textbf{9}, 263 (1994).

\bibitem{Jose85} Jorge V.\ Jos\'{e}, G.\ Monsivais, and J.\ Flores, Phys. Rev.\
    B \textbf{31}, 6906 (1985).

\bibitem{Diaz05} E.\ D\'{\i}az, A.\ Rodr\'{\i}guez, F.\ Dom\'{\i}nguez-Adame,
    and V.\ A.\ Malyshev, Europhys. Lett. \textbf{72}, 1018 (2005).

\bibitem{Dunlap86} D.\ H.\ Dunlap and V.\ M.\ Kenkre, Phys.\ Rev.\ B
    \textbf{34}, 3625 (1986).

\bibitem{Heijs04} D.\ J.\ Heijs, V.\ M.\ Malyshev, and J.\ Knoester, J.\ Chem.\
    Phys.\ \textbf{121}, 4884 (2004).

\bibitem{Malyshev95} V.\ Malyshev and P.\ Moreno, Phys.\ Rev.\ B \textbf{51},
    14587 (1995); A.\ V.\ Malyshev and A.\ V.\ Malyshev, Phys.\ Rev.\ B
    \textbf{63}, 195111 (2001).

\bibitem{Malyshev:unpublished} A.\ V.\ Malyshev and A.\ V.\ Malyshev,
    unpublished.

\bibitem{Fidder91} H.\ Fidder, J.\ Knoester, and D.\ A.\ Wiersma, J.\ Chem.\
    Phys.\ \textbf{95}, 7880 (1991).

\bibitem{Malyshev91} V.\ A.\ Malyshev, Opt.\ Spectrosk.\ \textbf{71}, 873 (1991)
    [Engl.\ Transl.: Opt.\ Spectrosc.\  \textbf{71}, 505 (1991)]; J.\
    Lumin.\  \textbf{55}, 225 (1993).

\bibitem{Knapp84} E.\ W.\ Knapp, Chem.\ Phys.\ Lett.\ \textbf{85}, 73 (1984).

\bibitem{Malyshev99} V.\ A.\ Malyshev, A.\ Rodr\'{\i}guez, and F.\
    Dom\'{\i}nguez-Adame, Phys.\ Rev.\ B \textbf{60}, 14\,140 (1999);
    F.\ Dom\'{\i}nguez-Adame and V.\ A.\ Malyshev, J.\ Lumin.\ \textbf{83-84},
    61 (1999).

\bibitem{Malyshev06} V.\ A.\ Malyshev and F.\ Dom\'{\i}nguez-Adame
    (in preparation).

\end{thebibliography}
\end{document}